\def\PR{{Phys.~Rev.~}}
\def\PRL{{ Phys.~Rev.~Lett.~}}
\def\PRB{{ Phys.~Rev.~B~}}

\def\etal{{\it et al.}}

\def\be{\begin {equation}}
\def\ee{\end {equation}}
\def\ber{\begin {eqnarray}}
\def\eer{\end {eqnarray}}
\def\bers{\begin {eqnarray*}}
\def\eers{\end {eqnarray*}}

\makeatletter

\newcommand{\Rmnum}[1]{\expandafter\@slowromancap\romannumeral #1@}
\makeatother

\makeatletter
\newcommand*\env@matrix[1][*\c@MaxMatrixCols c]{%
  \hskip -\arraycolsep
  \let\@ifnextchar\new@ifnextchar
  \array{#1}}
\makeatother

\documentclass[aps,prb,showpacs,superscriptaddress,a4paper,twocolumn]{revtex4-1}

\usepackage{color}

\usepackage{graphicx}

\begin {document}

\author{Banasree Sadhukhan}\email{banasree.rs@presiuniv.ac.in}
\affiliation{Department of Physics, Presidency University, 86/1College Street, Kolkata 700073, India}
\author{Prashant Singh}\email{prashant@ameslab.gov}
\affiliation{Ames Laboratory, U.S. Department of Energy, Iowa State University, Ames, Iowa 50011-3020, USA}
\author{Arabinda Nayak}\email{arabinda.physics@presiuniv.ac.in}
\affiliation{Department of Physics, Presidency University, 86/1College Street, Kolkata 700073, India}
\author{Sujoy Datta}
\affiliation{University of Calcutta, Physics Department, Acharya Prafulla Chandra Road, Kolkata, India}
\author{Duane D. Johnson}\email{ddj@iastate.edu}
\affiliation{Ames Laboratory, U.S. Department of Energy, Iowa State University, Ames, Iowa 50011-3020, USA}
\affiliation{Materials Science \& Engineering, Iowa State University, Ames, Iowa 50011-2300, USA}
\author{Abhijit Mookerjee}\email{abhijit@bose.res.in}
\affiliation{Professor Emeritus, S.N. Bose National Center for Basic Sciences, JD III, Salt Lake, Kolkata 700098, India}

\title{\bf Band-gap tuning and optical response of two-dimensional Si$_{x}$C$_{1-x}$: A first-principles real space study of disordered 2D materials.}
\date{\today}
 
\begin{abstract}
We present a real-space  formulation for calculating the electronic structure and optical conductivity of such random alloys based on the Kubo-Greenwood formalism interfaced with the augmented space recursion (ASR) [A. Mookerjee, J. Phys. C: Solid State Phys. {\bf 6}, 1340 (1973)] formulated with the Tight-binding Linear Muffin-tin Orbitals (TB-LMTO) basis with van Leeuwen-Baerends corrected exchange (vLB) [Singh \etal, Phys. Rev B {\bf 93}, 085204, (2016)]. This approach has been used to quantitatively analyze the effect of chemical disorder on the configuration averaged electronic properties and optical response of 2D honeycomb siliphene Si$_{x}$C$_{1-x}$ beyond the usual Dirac-cone approximation. We predicted the quantitative effect of disorder on both the electronic-structure and optical response over a wide energy range, and the results discussed in the light of the available experimental and other theoretical data. Our proposed formalism may open up a facile way for planned band gap engineering in opto-electronic applications.
\end{abstract}

\date{\today}

\maketitle

\section {Introduction}
{\par}The discovery of graphene and related two and quasi-two dimensional (2D) materials has accelerated  interest towards the fabrication of nanometer-scaled two-dimensional materials. This is because of the high mobility of their electrons and potential use in both mesoscopic research, materials engineering and nanodevices.\cite{geim,butl,novo1,novo2} However, the zero band gap of Graphene limits its applications in opto-electronic devices. Thereafter several attempts have been made to increase the band gap, e.g., using bilayer graphene with non-equivalent top and bottom layers,\cite{kats,zhang} or by surface doping and electric gating effect.\cite{kaplan} Another effective technique is the hydrogenation of graphene. A maximum band gap  of 0.77 eV has been  reached with 54\% of hydrogen coverage.\cite{balog,elias} Giovannetti~\etal  \cite{giovan} have also reported  substrate-induced band gap in graphene on hexagonal boron nitride. However a large band gap satisfying  commercial requirements of light-emitting diodes (LEDs) or solar cells is still not available.



{\par}In search of new material, recently, focus has been changed from graphene to other graphene$-$like 2D materials, with the aim of overcoming the shortage of graphene and broadening its range of applications.\cite{10,11} Such 2D layered materials can also be produced using parent 3D bulk materials.\cite{12} Among these silicene, MoS$_2$, germanine, etc. have been synthesized successfully and have been found to exhibit new physical properties.\cite{liu,tabert,qiu} In this work, we focus on 2D-layered structures of Siliphene, which can be understood in terms of graphene (2D-C) or silicene (2D-Si) doped with `Si' or `C', respectively, which is another Group IV binary compound displaying interesting properties. Since wurtzite silicon-carbide structure is graphitic in nature and theoretical investigations indicate that a phase transformation from graphinic to graphite-like structure is possible,\cite{26,27,28} and can surely be a good candidate with larger band-gap.

{\par}Earlier, first-principles calculations reported stable 2D-Si buckled honeycomb structures and their charge carriers also behave like massless Dirac fermions and their $\bf \pi$ and $\bf \pi$* bands cross linearly at the Fermi level. However, unlike 2D-Si, the mono-layers SiC have stable 2D planar honeycomb structures, which may be attributed to the strong ${\it p}$-bonding through the perpendicular ${\it p_z}$ orbitals,\cite{free,wu,hsueh} Previously some 2D silicon-carbon binary compounds including hexagonal rings have been proposed, such as hexagonal-SiC,\cite{lin1,lin2} tetragonal-SiC,\cite{chao}, g-SiC$_2$,\cite{zhou1} Pt-SiC$_2$,\cite{zhou2} SiC$_3$,\cite{ding} which shows interesting properties such as a large direct band gap, improved photoluminescence, and high-power conversion efficiency.\cite{free,wu,hsueh} Because the honeycomb structure is common to both C and Si, experimentally stable 2D-SiC in honeycomb structure has been synthesized successfully.  Recent progress in the fabrication of ultra thin layered 2D-Si$_{x}$C$_{1-x}$ down to a monolayer (0.5-1.5nm)\cite{hern,stan,stan1,lin} makes it an emerging semiconductor attractive to both fundamental research and practical applications. So, in this work, we emphasize on understanding electronic-structure and optical properties of the semiconducting 2D-Si$_{x}$C$_{1-x}$ material both qualitative and quantitative aspect. The choice of the system is based on the existing experiments for direct comparison and validation of our theoretical approach.

 


We have arranged the paper as follows. In Sec.~II, we describe the computational methodology. Results are discussed in Sec.~III.  When possible, comparisons to previous theoretical and experimental results are made. We summarized our results in Sec.~IV.


\begin{figure}[t]
\centering
\includegraphics[scale=0.19]{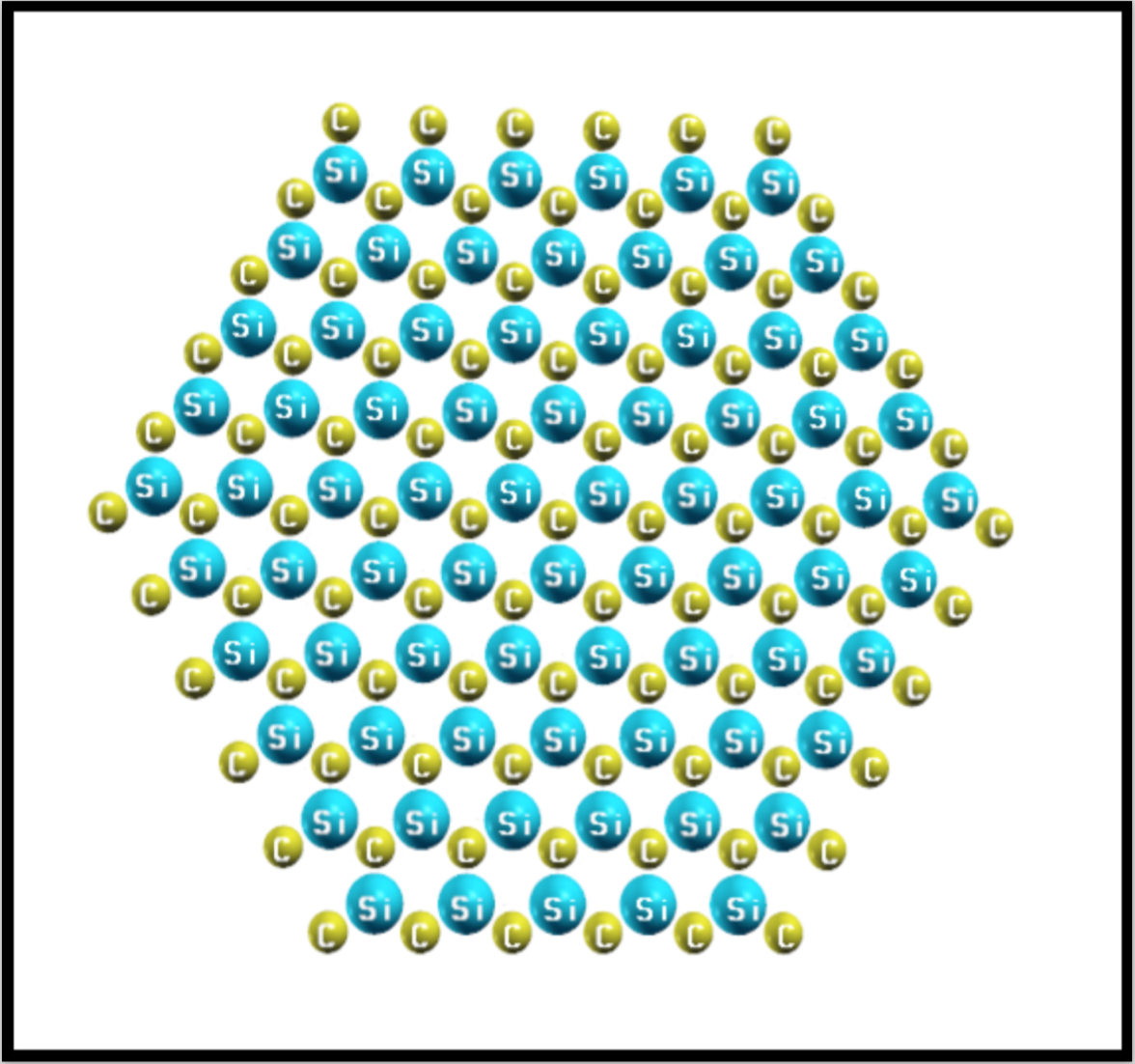}
\includegraphics[scale=0.2]{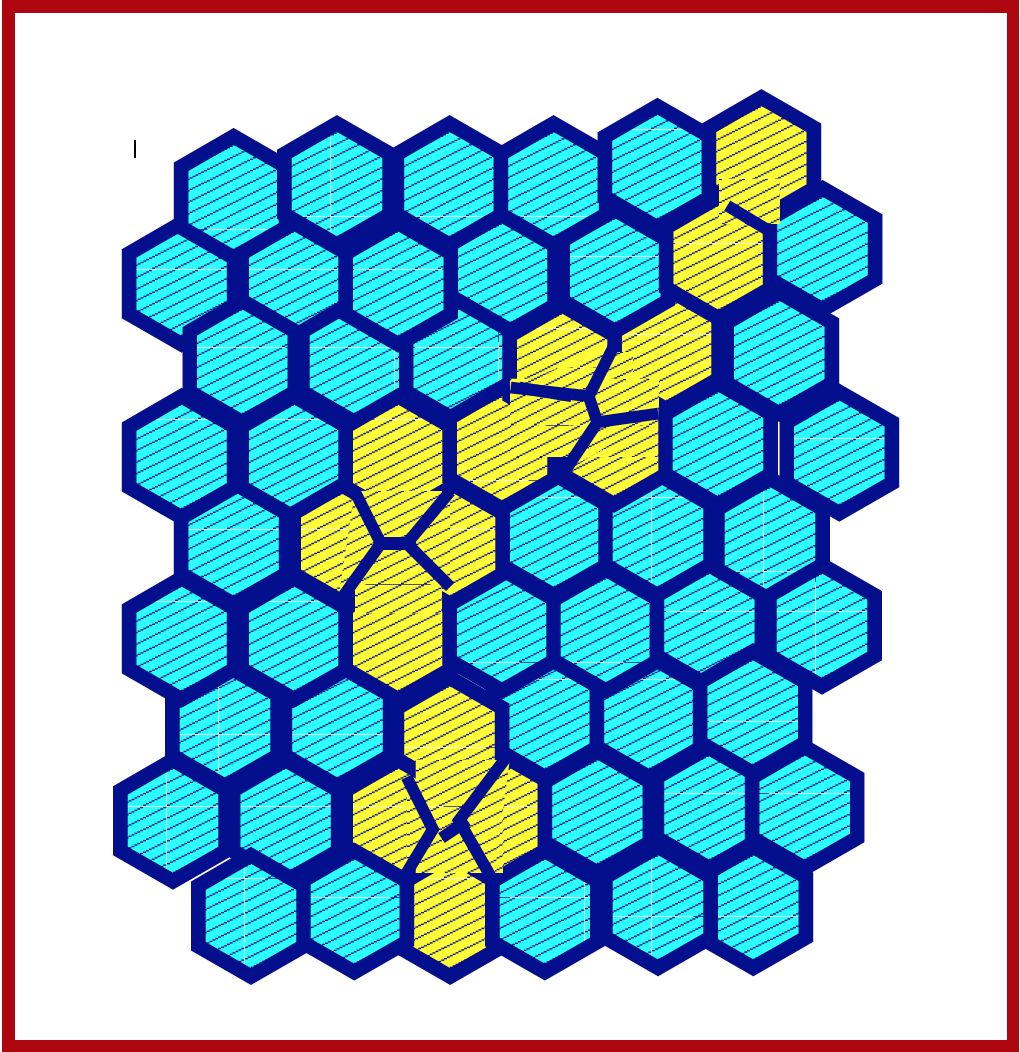}
\caption{(Color Online) (Left panel) Siliphene structure with atomic sizes of Si atoms much larger than those
for C. (Right panel) The formation of extended randomly placed chains of  Stone-Wales defects in graphene due to chemical disorder between Si and C on the hexagonal lattice.}
\label{tg} 
\end{figure}
 
\section{Computational details}

We treat random-disorder in alloys using augmented space recursion (ASR) technique.\cite{asr1,asr2,asr3,asr4,asr5,asr6,asr7} The ASR allows us to go beyond the local mean field theories like the coherent potential approximation and describe extended disorder like short-ranged clustering or ordering as well as long-ranged disorder like randomly placed Stone-Wales defect chains forming in a graphene background (left panel-Fig.\ref{tg}). The ASR can deal both chemical and local structural disorder.\cite{SM}

The ASR method is a real-space recursion technique for treating random disorder of alloys. The recursion\cite{viswa,term} needs a countable basis $\{|n>\} \in Z^d $, which is provided by the tight-binding linearized muffin-tin orbital (TB-LMTO) method.\cite{TBLMTO} Here, `n' is the basis and $\vec{R}_{n}$ represents atomic positions.\cite{heine, hhk,hay,hay2,hn,hn2} Since the ASR deals with the  configuration fluctuations in real-space, e.g. for binary randomness $\Pi^\otimes\ Z^2_n$, thus Bloch's theorem plays no role.
 
In this work, we use local-density approximation (LDA) modified with van Leeuwen-Baerends (vLB) \cite{VLB} correction to exchange part.\cite{singh2013,singh2016} The correction give a substantial improvement in the band gaps over the LDA. \cite{singh2013, singh2016} We use tight-binding linear muffin-tin (TB-LMTO)-vLB approach,\cite{TBLMTO, VLB, singh2013,singh2016} and combine it with augmented space recursion method \cite{asr1} to carry out configuration averaging in semiconducting random disordered systems. We use LDA correlation energy parameterized by van Barth and Hedin.\cite{vBH}  

All calculations were done self-consistently and non-relativistically for given experimental geometry till the ``averaged relative error'' between the converged final and the previous iteration charge density and energy is reached, i.e. 10$^{-5}$ and 10$^{-4}$ eV/atom respectively. We use the tetrahedron method for k-space integration, and Anderson mixing to facilitate convergence. In TB-LMTO-vLB core states are treated as atomic-like in a frozen-core approximation. Energetically higher-lying valence states are addressed in the self-consistent calculations of the effective crystal potential, which is constructed by overlapping Wigner-Seitz spheres for each atom in the unit cell.

To assess the optical conductivity of disorder system, we use Kubo-Greenwood approach (KGA) \cite{kubo,greenwood} with ASR technique.\cite{asr1,moshi, singh2016} The KGA-ASR technique is used for systematic study of band gap tuning and configuration averaged optical responses in random alloys, which can be a unique formalism for calculating optical response of semiconducting alloys, both bulk and finite size.  

The detailed discussion of mathematical approach has been provided in the Appendix.

\section{Results and Discussion.}
{\par}We performed first-principles calculations on Si$_{x}$C$_{1-x}$ ( 0 $<\ x\ \le $ 0.6) and compared our predictions with some existing experimental and theoretical results.\cite{azadeh,lin} We chose homogeneously disordered binary 2D-siliphene (Si$_{x}$C$_{1-x}$), specifically because of several advantages over graphene.\cite{nako,vogt} One of them is a better tunability of the band gap, very important in designing optoelectronic nano-devices. There has been disagreement between different theoretical and experimental works on band gap engineering in honeycomb 2D-Siliphene. A quick review indicates that the main cause for these disagreements, at least among the theoretical studies, has been the dependence of the results on the underlying models and approximations, e.g. the starting structure, the type of exchange-correlation, the electronic structure method used and how disorder is taken into account, include effects of configuration fluctuations of the immediate neighborhood. These are the focus points of our investigation, as described earlier. In all calculations, we consider the homogenous disorder in Si$_{x}$C$_{1-x}$.\cite{ding,shi2015}  In their recent work, Ding \etal \cite{ding} and Shi \etal \cite{shi2015}, has shown the existence of homogenous disorder mostly for Si-concentration $\le$50 at.\% in 2D Si$_{x}$C$_{1-x}$. 

Experimentally, graphene shows zero band gap at a lattice constant of $a$ = 2.46 $\AA$, while siliphene has a 1.90 meV gap with $a$ = 3.86 $\AA$. \cite{NJSBKMG2005} The energy difference of 10.66~eV ( E$_{2s}-$E$_{2p}$) for graphene, and 5.66~eV (E$_{3s}-$E$_{3p}$) for silicene indicate predominantly $sp^{2} + p_{z}$ hybridization due to presence of graphene valence cloud. Thus graphene is more stable as  a flat surface, while the valence cloud in siliphene has the possibility of $sp^3$ hybridization leading to local corrugations and buckling. In our study we have not included buckling. 

\begin{figure}[t]
\centering
\includegraphics[scale=0.35]{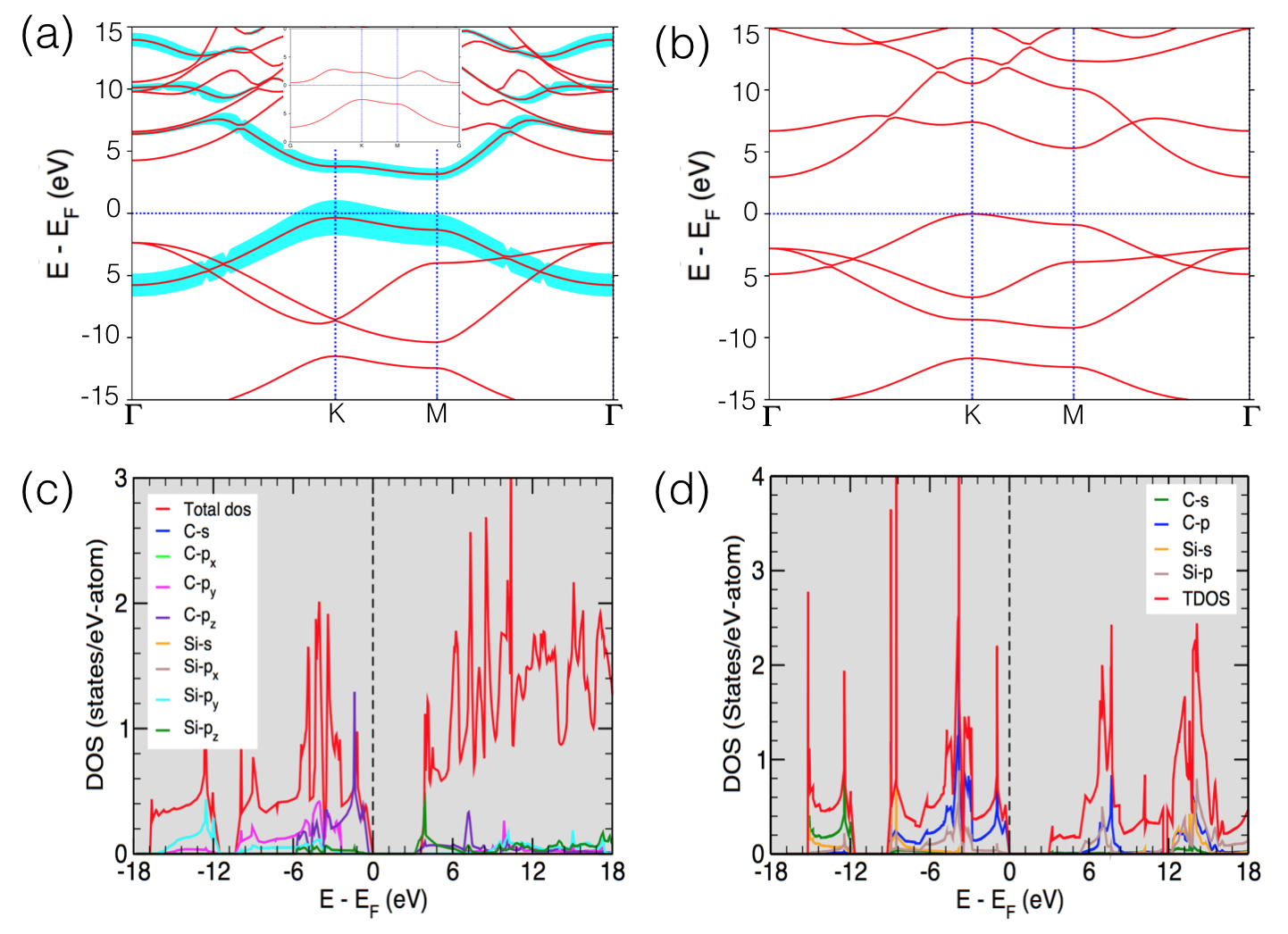}
\caption{(Color Online) Plots of NMTO (left panel) (a) band structure, and (c) DOS, TB-LMTO-vLB calculated (b) band structure, and (d)  DOS (right panel) at lattice constant a=2.66 $\AA$ of 2D-Si$_{0.50}$C$_{0.50}$. The shaded band is for $\it{p_{z}}$ orbital without down-folding while the bands in inset are the down-folded $\it{p_{z}}$ orbitals. The magnitude of the band gap from the vLB-LMTO is in greater agreement with experiment.}
\label{fig2} 
\end{figure}

{\par}The experiments of Lin~\etal \cite{lin} led to lattice constants ranging from 2.4 $\AA$ to 2.8 $\AA$ of 50-50 2D-Siliphene using high resolution transmission electron microscope. GIXRD analysis confirmed $a$ = 2.66 $\AA$. The two earlier theoretical approaches were by  Bekaroglu~\etal  \cite{bekaroglu} and the Azadeh~\etal \cite{azadeh} Both  used $\vec{k}$-space DFT approaches.\cite{VASP,SIESTA, WIEN2k} This  restricted their  study to ordered stoichiometric compositions in strict accordance with Bloch's theorem, ignoring the effects of disorder. For 2D-Si$_{x}$C$_{1-x}$ (x=0.5), Bekaroglu~\etal~\cite{bekaroglu} used generalized gradient approximation (GGA) within projector augmented wave approach \cite{PAW} and proposed theoretical lattice constant  $a\ \sim$ 3.094 $\AA$ nearly equal to that for  Wurtzite SiC ($a$ = 3.08 $\AA$). On the other hand, Azadeh~\etal \cite{azadeh} introduce an interesting in-crystal-site doping or combination stoichiometric method to generate some `semi-random' compositions and studied them using the  WIEN2K code.\cite{WIEN2k}  

In Table~\ref{tab1}, we compare the results calculated from vLB-LMTO with other theoretical methods and measurements. The energy minimized lattice constant of 2D-SiC using vLB-LMTO is in good agreement with the measured a=2.6$\pm$0.2 $\AA$. However, we found good agreement between our calculations and the experiments but theoretical results differ almost by $\sim$15\%.  The main source of disagreement of other theoretical approaches with vLB-LMTO arise from the use of exchange-correlation potential. We use vLB+LDA exchange-correlation,\cite{singh2016} which betters both r$\rightarrow$0 zero and $1/r$ asymptotic limits. Also, vLB cancels self-interaction better than other density based functionals.\cite{singh2016}

In Fig.~\ref{fig2}(a)\&(c), we plot Nth-order muffin-tin orbital (NMTO)\cite{PRB2K} method calculated (a) band-structure, and (c) density of states. The maximum DOS contribution comes from the ${\sl p_{z}}$ orbital with indirect (direct) band gap of 3.59 (3.90) eV along K$-$M (K$-$K) symmetry direction. However, the vLB-corrected LDA\cite{singh2016} provides an accurate band structure and density of states, shown in Fig.\ref{fig2} (b)\&(d), calculated at equilibrium lattice constant of 2.66 $\AA$. The vLB+LDA provides indirect band gap of 2.96 eV along K$-\Gamma$ symmetry direction of the Brilluoin zone. Siliphene (Si$_{x}$C$_{1-x}$,  x=0.5), theoretical equivalent of graphene with silicon doping, is also supposed to be gap less. However, the honeycomb lattice of randomly but homogeneously distributed silicon atoms in symmetric semi-metallic graphene opens up a gap, and shows semi-conducting behavior. This way, the silicon doping allows us to tune the band gap in the Si$_{x}$C$_{1-x}$.

\begin{figure}[b]
\centering
\includegraphics[scale=0.35]{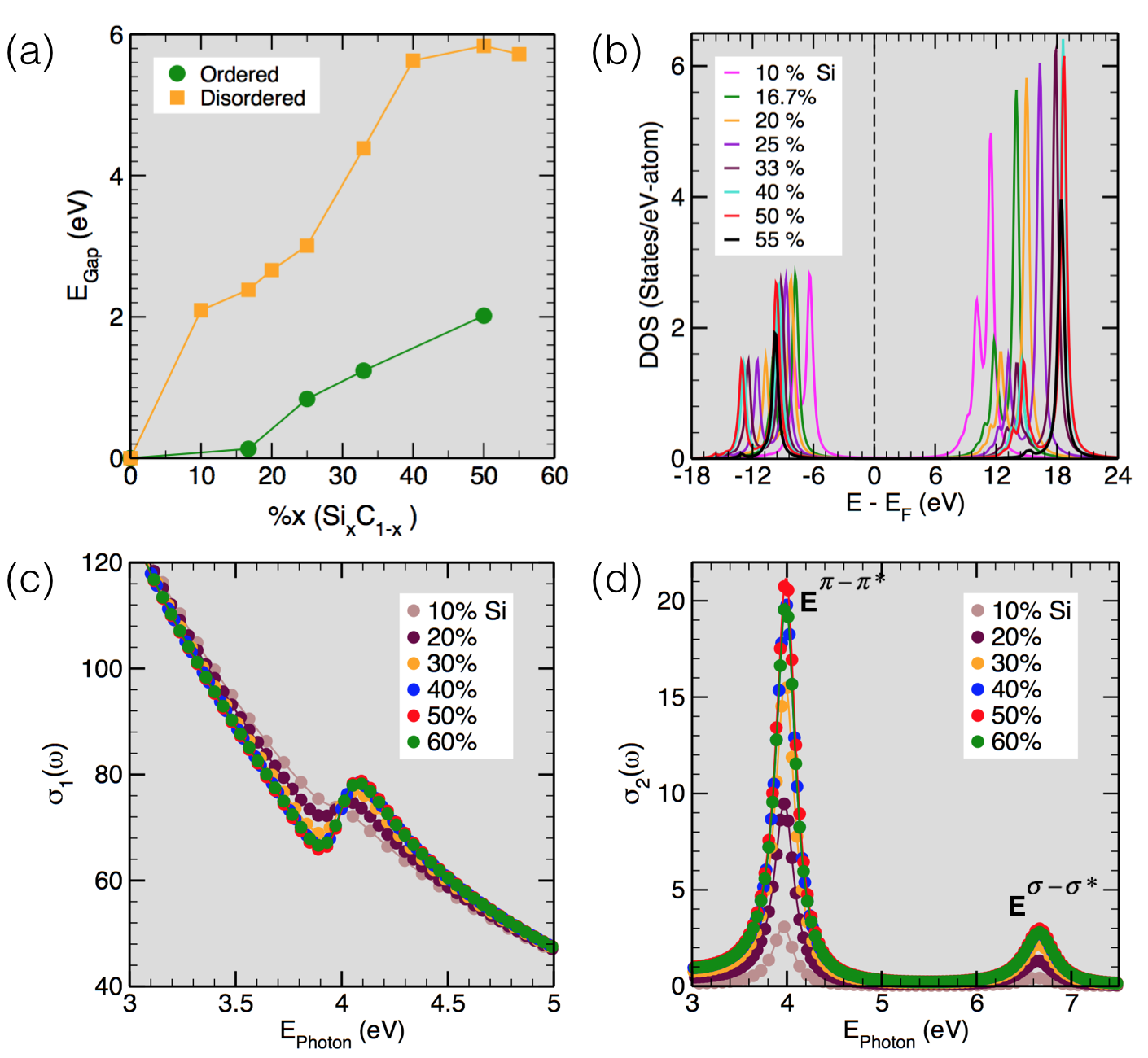} 
\caption{(Color Online) We plot (a) Band gap variation (order and disordered Si$_{x}$C$_{1-x}$), (b) total density of states, (c) real part, and (d) imaginary part of optical conductivity of disordered Si$_{x}$C$_{1-x}$ at \%Si dependent lattice constants (see Fig.~\ref{fig1A}). The band gap variation for ordered from Azadeh~\etal,\cite{azadeh} while for disordered case is from TB-LMTO-ASR (using Cambridge recursion library). The maximum in band gap is observed at 50\%Si, i.e., Siliphene (Si$_{0.50}$C$_{0.50}$). The optical conductivity peak at 3.96 eV, in (c) and (d), is in agreement with experimentally observed peak at 3.33 eV.\cite{lin}}
\label{fig3}
\end{figure}

We show, in Fig.~\ref{fig3}(a)\&(b), the band gap and TDOS (total density of states) variation with changing lattice constant for given \%Si ( see Fig.~\ref{fig1A} in appendix, composition dependent  lattice parameters). The band gap clearly increases, as shown in Table~\ref{tab2} and Fig.~\ref{fig3}(a), with increasing \%Si and reaches maximum at 50\%, and  further increasing \%Si reduces the band gap. The large atomic size of Si enhances the interaction strength, and also hybridization between the p$_z$ orbital for both Si and C. While the $ \bf \sigma$-band remains unchanged contrary to $\bf \pi$-band. 

\begin{table*}
\centering
\begin{tabular}{c|c||c|c||c|c||c|c||c|c||c| c}\hline\hline
\multicolumn{10}{c||}{Bekaroglu \etal\cite{bekaroglu} (x=0.5)} & \multicolumn{2}{c}{Azadeh \etal (x=0.5)}\\ \hline
\multicolumn{2}{c||}{PAW-LDA}& \multicolumn{2}{c||}{PAW-GGA}& \multicolumn{2}{c||}{US-LDA}& \multicolumn{2}{c||}{US-GGA}& \multicolumn{2}{c||}{LDA-GW}& \multicolumn{2}{c}{VASP/SIESTA}\\ \hline
$a$ (\AA) & E$_{g}$ (eV) & $a$ (\AA) & E$_{g}$ (eV) & $a$ (\AA) & E$_{g}$ (eV) & $a$ (\AA) & E$_{g}$ (eV) & $a$ (\AA) & E$_{g}$ (eV) & $a$ (\AA) & E$_{g}$ (eV)\\ \hline
3.07 & 2.51 & 3.09& 2.53 & 3.05 & 2.53 & 3.08 & 2.54 & {--}& 3.90 & 2.41  & 2.06  \\
\hline\hline
\end{tabular}
\vskip 0.5cm
\begin{tabular}{c|c||c|c||c|c||c|c||c| c||c| c}\hline\hline
\multicolumn{4}{c||}{Lin \etal (x=0.5) Experiment} & \multicolumn{6}{c||}{Our Work (x=0.5)}& 
\multicolumn{2}{c}{Our Work (x=0.5)}\\ \hline
\multicolumn{2}{c||}{TEM}& \multicolumn{2}{c||}{GIXRD}& \multicolumn{2}{c||}{\enskip LDA-LMTO\enskip}& \multicolumn{2}{c||}{vLB-LMTO}& \multicolumn{2}{c||}{vLB-ASR}& \multicolumn{2}{c}{NMTO}\\ \hline
$a$ (\AA) & E$_{g}$ (eV) & $a$ (\AA) & E$_{g}$ (eV) & $a$ (\AA) & E$_{g}$ (eV) & $a$ (\AA) & E$_{g}$ (eV) & $a$ (\AA) & E$_{g}$ (eV)&  $a$ (\AA) &\qquad E$_{g}$ (eV)\enskip  \\ \hline

2.60$\pm$0.2 & 2.6-2.9 & 2.66 & 2.95 & 2.66 & 2.96 & 2.66 & 3.49 & 2.75 & 3.60 & 2.66& 3.90  \\
\hline\hline
\end{tabular}
\caption{The calculated lattice constant and band gap for 50$\% $ Siliphene using vLB-LMTO and its comparison with other
theoretical approaches and experiments. Good agreement between vLB calculated lattice constants are in good agreement with experiments.\cite{lin} Also vLB shows better band-gap with respect to other methods due to reduced self-interaction, and the better asymptotic treatment.\cite{singh2016}} 
\label{tab1}
\end{table*}

\begin{table}[b]
\begin{tabular}{cccccc}\hline\hline 
 &&&\multicolumn{3}{c}{Band Gap (eV)}\\
 \%x &&& {Azadeh~{\etal}} &&  {This work}\\ 
 \ &&& {(ordered)}      &&  {(disordered)}\\ \hline
{  0.10} &&&{   -- }      && { 2.094} \\
{  0.17 }&&&{ 0.13}    && { 2.383} \\
{  0.20}&&&{   --  }      && { 2.662} \\
{  0.25}&&&{ 0.838 }  && { 3.008} \\
{  0.33}&&&{ 1.237 }  && { 4.387} \\
{  0.40}&&&{   --     }  && { 5.628} \\
{  0.50 }&&&{ 2.061} && { 5.835} \\
{  0.55 }&&&{   --  }   && { 5.719 } \\
\hline
\end{tabular}
\caption{The band gaps calculated for disorder 2D-Si$_{x}$C$_{1-x}$ using LDA-vLB exchange-correlation potential within augmented space formalism. We compare our calculations with the work of Azadeh~\etal \cite{azadeh} done at some specific compositions of semi-random alloys.}
\label{tab2}
\end{table}

{\par}Repulsive effect between the valence band and the conduction band increases with increasing Si content and reduced C$-$Si bond length. The small bond lengths enhances in-plane $\pi-\pi$ repulsion, which helps opening up the band gap. The anti-bonding states moves towards a higher energy level due to increased kinetic energy and repulsion effect of $\it{p}-$electrons with increasing \%Si. Also, delocalization effect due to orbital overlapping between C and Si plays important role in band gap opening with increasing disorder strength. The reduced bond-length enhances delocalization, which broadens both the valence and conduction bands and decreases the band gaps.  Assuming silicene as the end point of Si$_{x}$C$_{1-x}$) with x=1.0, one would expect the downturn of the mentioned trend beyond 50\% Si. For graphene the effective mass at Dirac point is at zero, while in other graphene-like materials with finite band gap has larger effective mass and reduced mobility. This establishes that increasing strength of chemical disorder increases the band gap up to certain \%Si, in agreement with the previous reports.\cite{azadeh}

 {\par}Many-body interactions are more significant in 2D materials than in their bulk counterparts.\cite{kin1,kin2,kin3} This reflects the intrinsic enhancement of the importance of Coulomb interactions in 2D materials and their reduced screening. We note that the theoretical exciton binding energy in bulk 2H-SiC is only 0.1 eV, whereas it is 1.17 eV for 2D honeycomb SiC.\cite{hsueh} This shows that the reduced dimensionality of a SiC sheet confines the quasi-particles. This significantly enhances the overlap between the electron and the hole wave functions and hence the electron-hole (e$-$h) interaction. The presence of the vacuum region reduces the screening and hence provides an extra contribution to the large excitonic effect in the SiC sheet in addition to this quantum confinement,. The role of e$-$h interactions is relevant for the optical response. The repulsive e$-$e interactions shifts the transition peak upwards in energy, and the attractive e$-$h interactions shifts it downwards.\cite{AKK2008,KOER2010,KTB2011} The optical dielectric function of the 2D-SiC sheet is highly anisotropic. Therefore, many-body interaction effects in the $\pi$-band associated with the p$_z$ orbitals, which extend into the vacuum region from the sheet, would be less screened. Consequently, the $\pi$-band would have a larger quasi-particle corrections. On the other hand, because the in-plane $\sigma$-bonds are mainly confined to the 2D-SiC sheet, screening effects on the $\sigma$-bands are more significant, and hence quasi-particle corrections are smaller. 

The real and imaginary part of optical conductivity at different Si content are shown in Fig.~\ref{fig3} (c) and (d). We found an optical conductivity peak at 3.96 eV in the low-energy region, however, at higher energies another peak is found at 6.63 eV. Clearly, increasing \%Si increases the band gap (till 50\%Si), and optical conductivity peaks exhibit noticeable broadening. The small shift in 55\%Si peak (goes below 50\%Si) in Fig.~\ref{fig3}(b), which is in agreement with band gap change above 50\%Si.  The optical transition peak implicitly depends on the band gap, and on the factor $\left\langle \tilde{j}(t)\  \tilde{j}(0) \right\rangle$ $\delta$ (E$_f$ - E$_i$ + $\hbar\omega$). The transition peak is shadowed with increasing disorder strength (x) in 2D-Si$_{1-x}$C$_{x}$. The reported transition peak by Hsueh~\etal \cite{hsueh} from GW-Bethe-Salpeter equation for single 2D-SiC sheet have similar magnitude, see Table.~\ref{tab3}. 

In the low-energy range (2$-$5 eV), the inter-band optical transitions involve mainly the $\pi$-bands, however, at higher energies, the optical absorption peaks between 5$-$11 eV are associated with the inter-band transitions involving $\sigma$-bands. The first prominent peak, located at 3.96 eV arises due to the excitation between the $\pi$ and the $\pi^*$ states and the pronounced optical peak peak at 6.63 eV is mainly due to the excitation between the $\sigma$ and the $\sigma^*$ states. The optical conductivity peak at 3.96 eV is in good agreement with experimental peak at 3.33 eV  as reported by Lin~\etal, \cite{lin} see Table.~\ref{tab3}. The small deviation from their data arises due to multi-layers structures. They did their experiment in 2D-Si$_{x}$C$_{1-x}$ nano-sheets of thickness $<$ 10 nm, and the interactions between substrates and samples was taken into consideration in the high-resolution transmission electron microscopic measurements.

\begin{table}[h]  
\begin{tabular}{ccccccccccc}\hline \hline
{2D Systems  }&{Theory }&{ Experiment }\\ 
&{  (eV)  }&{  (eV) }\\ \hline 
  &{  }&{   }\\
{Monolayer }&{  3.96 $\pi$ $\rightarrow$ $\pi*$ }&{  --  }\\
{ Si$_x$C$_{1-x}$ }&{\small 6.63 $\sigma$ $\rightarrow$ $\sigma*$ }&{   }\\
&{   [ vLB-ASR ] }&{    }\\
&{  }&{   }\\
{Single SiC sheet }&{  3.25 $\pi$ $\rightarrow$ $\pi*$ }&{  --  }\\
&{  5.83 $\sigma$ $\rightarrow$ $\sigma*$ }&{   }\\
&{   [ GW+BSE \cite{hsueh}] }&{    }\\
&{  }&{   }\\
{ Single SiC sheet }&{\small 4.42 $\pi$ $\rightarrow$ $\pi*$ }&{  --  }\\
&{  6.20 $\sigma$ $\rightarrow$ $\sigma*$ }&{   }\\
&{   [ GW+RPA \cite{hsueh}] }&{    }\\
&{  }&{   }\\
{Ultra thin SiC nano-sheets }&{  - }&{  3.33 $\pi$ $\rightarrow$ $\pi*$} \\
{($<$10 nm) }&{   }&{  \cite{lin} }\\
&{  }&{   }\\
\hline \hline
\end{tabular}
\caption {(Color online) The calculated optical conductivity peak of disordered Si$_{x}$C$_{1-x}$ (at x=0.5) is presented and compared with theory \cite{hsueh} and experiment.\cite{lin}}
\label{tab3}
\end{table} 

The results shown in Table.~\ref{tab1} followed by subsequent discussions indicates that the calculated electronic structure depends upon the model chosen, the approximations introduced and the calculational methodology followed. The predictions of our proposed model are in good agreement with the experiments. The band gap and optical conductivity calculated at equilibrium lattice parameters are closest to the experiments. The effect of going beyond single-site is visible in calculated physical properties from ASR used with vLB.

 \section{CONCLUSION.}
In summary, we perform DFT calculations with a LDA-vLB exchange-correlation function to calculate the optical conductivity of 2D material, real and imaginary parts, of an isolated single-atom-thick layer consisting of a group-IV honeycomb crystal. The LDA and PBE functionals failed to  capture important feature in the visible light region, which is important for nanoplasmonic applications. The approach, we introduce reduces the degree of underestimation of the transition energy. Here, we focused our studies on disordered 2D-Si$_{x}$C$_{1-x}$,  which shows good agreement with experimental observations. We also found that silicon can be used to tune the band-gap in Si$_{x}$C$_{1-x}$, which in turn enhances the optical conductivity. The prediction of large optical response in our calculations shows that  2D-Si$_{x}$C$_{1-x}$ can be a potential candidate for the solar cell material. The proposed formalism opens up a facile way to band gap engineer material for optoelectronic application.

\section{ACKNOWLEDGMENT}
We are grateful to Prof. Andersen for giving us permission to take apart his LMTO-package and incorporate our new exchange functional and then coupling it with the Cambridge Recursion Package of Haydock and Nex. BS and AM thank R. Haydock and C.M.M. Nex for permission to use and modify the Cambridge Recursion Package and Yoshiro Nohara, Max Plank Institute, Stuttgart, Germany, for his kind permission to use TB-NMTO code developed by our group also. PS and DDJ support is from the U.S. Department of Energy (DOE), Office of Science, Basic Energy Sciences, Materials Science and Engineering Division; our research was performed at the Ames Laboratory, which is operated for the U.S. DOE by Iowa State University under Contract No. DE-AC02-07CH11358.
\vskip 0.5cm
\appendix*
\centerline{\bf APPENDIX}
In the Appendix we give a concise description of those points which we have introduced in this work. This
would give the reader a clear picture of our contribution.

 \subsection{ The van Leewen-Baerends Exchange.}
To address the band gap problem, we have proposed the use of spin-polarized van Leeuwen-Baerends (vLB) \cite{VLB} corrected exchange potential with a useful addition to satisfy the ionization potential theorem and to make the ionization energy and largest HOMO-LUMO difference agree in first-principle calculations. The asymptotic behavior of exchange was matched at the atomic sphere boundary (or local interstitial). The vLB exchange has earlier given excellent results in other applications.\cite{singh2013,singh2016} The vLB exchange-correlation  can be written as~:

\begin{equation}
V_{xc,\sigma}^{\rm model}(\vec{ r}) = \left[V_{x,\sigma}(\vec{ r}) + V_{x,\sigma}^{vLB}(\vec{ r})\right] + V_{c,\sigma}(\vec{ r})
\label{MODEL}
\end{equation}
where $V_{x,\sigma}(\bf r)$ and $V_{c,\sigma}(\vec{ r})$] are the standard LDA exchange and correlation  potentials and $V_{x,\sigma}^{\rm vLB}(\vec{ r})$ is the correction to LDA exchange. The suffix $\sigma$ represents the spin degree of freedom. Here, $V_{x,\sigma}^{vLB}(\vec{r})$ is 
\begin{equation}
V_{x,\sigma}^{\rm vLB} (\vec{r}) = -\beta\rho_{\sigma}^{1/3}\frac{x_{\sigma}^{2}}{1+3{\beta x_{\sigma}}{\sinh}^{-1}(x_{\sigma})},
\label{VLB}
\end{equation}
where  parameter $\beta=0.05$ was proposed originally by van Leeuwen and Baerends.\cite{VLB} Here, $x=|\nabla\rho(\vec{ r})|/{\rho^{4/3}(\vec{ r})}$ signifies the change in mean electronic distance provided density is slowly varying in given region and with strong dependence on gradient of local radius of the atomic sphere $R_{ASA}$.

The effective Kong-Sham potential becomes
\ber 
V_{\rm eff}(\vec{ r}) = V_{ext}(\vec{ r})+ V_{H}(\vec{ r}) + [V_{x,\sigma}(\vec{ r})+V_{x,\sigma}^{vLB}(\vec{ r})+V_{c,\sigma}(\vec{ r})]\phantom{XIX} ,
\label{KS}\nonumber\eer
where V$_{ext}(\vec{ r})$ is the external potential, V$_{H}(\vec{ r})$ is the electronic Hartree potential, V$_{x,\sigma}(\vec{ r})$ is LDA exchange,  V$_{c,\sigma}(\vec{ r})$ is LDA correlation, and V$_{x,\sigma}^{vLB}(\vec{ r})$ is the added spin-polarized  vLB-exchange. The iterative Kohn-Sham scheme  now has the effective potential constructed using new electronic density term
\ber \left\{-\frac{1}{2} \nabla^2+ V_{\rm eff}(\vec{ r})\right\} \phi_{i,\sigma}(\vec{ r}) = \epsilon_{i,\sigma}\phi_{i,\sigma}(\vec{ r})\phantom{XX XX}
\label{KS}\eer
We solve the Kohn-Sham  equation using the tight-binding linear muffin-tin orbital method with atomic sphere approximation (TB-LMTO-ASA)  \cite{TBLMTO} to obtain the exchange-correlation potential in the atomic-sphere. The solution is obtained iteratively to self-consistency.


\subsection{Lattice constant and Band-gap variation with composition in Si$_{x}$C$_{1-x}$:}

\begin{figure}[t]
\centering
\includegraphics[scale=0.4]{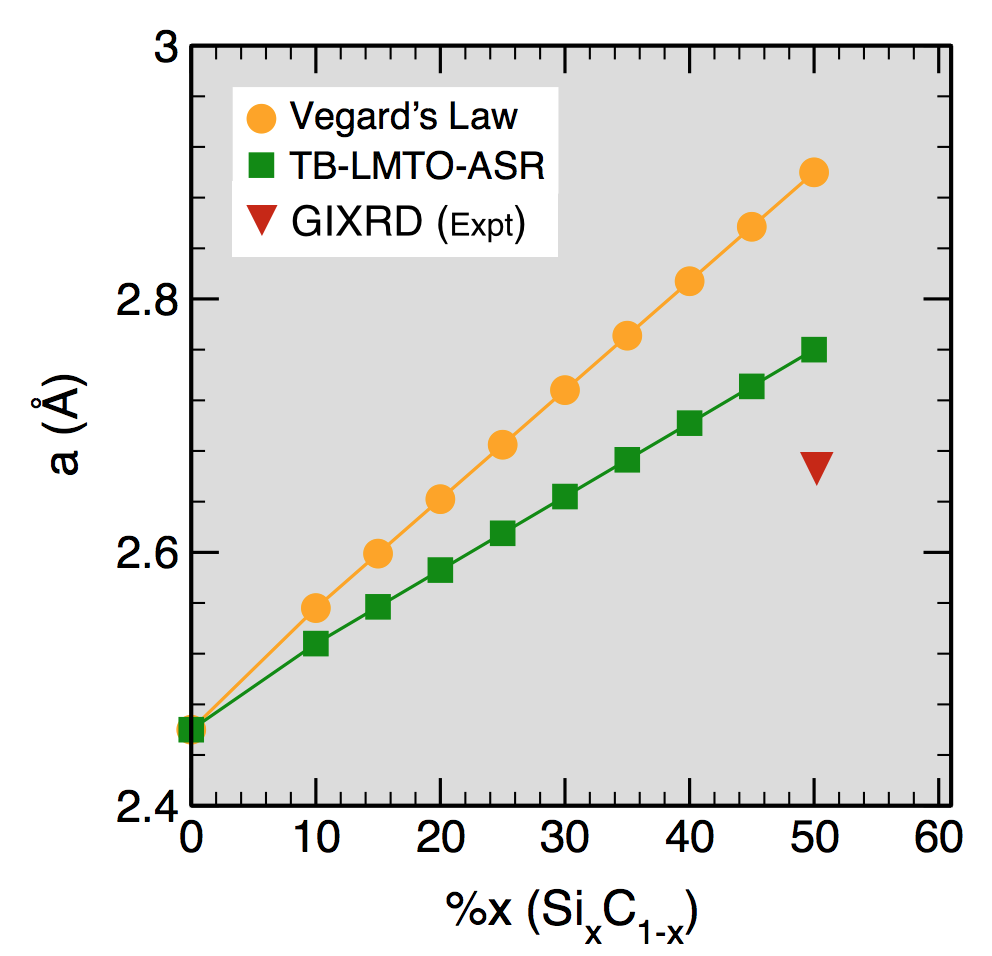} 
\caption{(Color Online) The variation of energy minimized lattice constant (a, \AA) in Si$_{x}$C$_{1-x}$ with alloy composition (0 $<$ x $<$ 60\%). Calculated `a' Vs \%x shows deviation from usual Vegard's law due to atomic size mismatch in `Si' (1.15 \AA) and `C' (0.77 \AA). The 'bowing-effect' in TB-LMTO-ASR (using vLB exchange-correlation potential) shows better agreement with experiment.\cite{lin}}
\label{fig1A} 
\end{figure}

We treat disorder using augmented space formalism, which includes the effect of near-neighbor environment (beyond single-site unlike CPA). \cite{asr1,asr2,asr3,asr4,asr5,asr6,asr7} The equilibrium lattice constant is calculated using vLB corrected exchange correlation potential within ASR formalism at different Si\%. The Vegard's Law usually works better for elements with similar sizes, while `C' (0.77) and `Si' (1.15) are of different atomic-sizes, both locally (in the intra-atom sp-hybridization) and globally (in inter-atomic bonding) in the solid. This atomic-size difference leads to `bowing effect' in Vegard's Law for 2D-SiC, see Fig.~\ref{fig1A}. 
  
\subsection{Real Space and Recursion.}
Keeping in mind our aim to develop a formalism without application of Bloch Theorem in any form, we have seamlessly 
 combined the modified LMTO with the Cambridge Recursion Library and the Augmented Space Recursion packages.  We propose this new TB-LMTO-vLB-ASR to be our technique to tackle disordered semi-conductors.

Since we shall be working in real space in Recursion, we express the Kubo response of the valence electrons to optical disturbance in a form so that the necessary correlation functions can also be expressed through a generalized recursion technique.\cite{viswa}

\be  \langle j^\mu(t)\rangle = \int_{-\infty}^\infty dt^\prime \sum_\nu \chi^{\mu\nu}(t-t^\prime) A^\nu(t^\prime) \ee

where
\[ \chi^{\mu\nu}(t-t^\prime) = \frac{i}{\hbar} \Theta(t-t^\prime) \langle G| [j^\mu(t),j^\nu(t^\prime)]|G\rangle
\]
where $|G\rangle$ is the ground state. In lattices with cubic symmetry, $\chi^{\mu\nu} = \chi \delta_{\mu\nu}$. The Fluctuation-Dissipation Theorem relates the imaginary part of the Laplace transform of the generalized susceptibility $\chi^{\prime\prime}(\omega)$ to the current-current correlation function~:

\begin{eqnarray*}
\chi^{\prime\prime}(\omega) & = & \frac{1}{2\hbar} \left[ \frac{\phantom{|}}{\phantom{|}}1-\exp(-\beta\hbar\omega)\right]\ S(\omega) \\ 
{\rm where} & & \\  
 S(\omega)& = &\int_0^\infty \ dt\ \exp(i\omega t)\ \langle G \vert j(t) j(0) \vert G\rangle
\end{eqnarray*}

The correlation function can be obtained most directly using the generalization of the recursion method. This has been described in detail by Viswanath and M\"uller  \cite{viswa,term}. Analogous to the standard Recursion for single particle propagators, we generate a new basis from a three term recurrence : 

\begin{enumerate}
\item[(i)] $ |f_{-1}\rangle = 0\enskip ;\enskip |f_0\rangle = j(0)| G\rangle$ 
\item[(ii)] We generate the rest of the basis recursively :
\[ |f_{m+1}\rangle = H|f_m\rangle - \alpha_m |f_m\rangle - \beta_m^2 |f_{m-1}\rangle \quad m=0,1,2\ldots \]
\item[(iii)] Mutual orthonormality leads to :
                \[ \alpha_m = \frac {\langle f_m|H|f_m\rangle}{\langle f_m|f_m\rangle} \quad \mbox{and} \quad 
\beta^2_m = \frac{\langle f_m|f_m\rangle}{\langle f_{m-1}|f_{m-1}\rangle} \]
\end{enumerate}
We can now expand any wave ket $|\psi\rangle$ in terms of the orthogonal basis just generated :
\[|\psi(t)\rangle \quad =\quad \sum_{m=0}^\infty \ C_m(t) |f_m\rangle 
\]

\be 
i \frac{d}{dt} C_m(t) = C_{m-1}(t) +\alpha_m C_m(t) + \beta_{m+1}^2 C_{m+1}
\label{eqn1}
\ee

with  $C_{-1}$=0 and $C_m(0)=\delta_{m,0}$.  Taking  Laplace transform we get,

\be 
(z-\alpha_m) C_m(z) + i \delta_{m,0} = C_{m-1}(z) +\beta^2_{m+1} C_{m+1}(z)
\ee

Recursive steps now give us :

\[C(z)=C_0(z) = \frac{-i}{\displaystyle z-\alpha_0-\frac{\beta_1^2}{\displaystyle z-\alpha_1-\frac{\beta_2^2}{\displaystyle z-\alpha_2-\ldots}}}\]

So that :  \be  S(\omega) = 2 \lim_{\delta\rightarrow 0}\ \Re e\ C(\omega + i\delta) \ee 
                                     
 \subsection{Disorder and the Augmented Space approach.}
 We model local disorder with associating with the tight-binding Hamiltonian elements a set of random parameters $\{n_m\}$. For binary alloying, for example, $n_m$ takes the values  0 and 1 with probabilities $x$ and $y$.

\begin{widetext}                                                         
\begin{eqnarray}[t]
 H & = & \sum_m \left(\epsilon_A n_m + \epsilon_B (1-n_m)\right) {\cal P}_m 
 + \sum_m \sum_{m'} \left\{ t_{AA} n_m n_{m'}
+ t_{BB} (1-n_m)(1-n_{m'}) \frac{\phantom{.}}{\phantom{.}} \ldots\right.\nonumber\\
& & + \left. t_{AB} \left\{n_m(1-n_{m'})+(1-n_m)n_{m'}\right\}\frac{\phantom{.}}{\phantom{.}}\right\} {\cal T}_{mm'} 
\end{eqnarray}
\end{widetext}

with the projection and transfer operators ${\cal P}_m$ and ${\cal T}_{mn}\ \in {\cal H}$.
Following the ideas of measurement theory, the prescription is to replace
the random parameters by the operator $\tilde{N}_m $ \cite{asr2} whose spectrum are the possible results
of measurement and whose spectral density is the probability of obtaining these results.
For example, if $n_m$ takes the values 0 and 1 with probabilities $x$ and $y$  then  $\tilde{N}_m $ is of rank 2 with a representation :

\[  \left( \begin{array}{cc} 
  x & \sqrt{xy}\\ 
  \sqrt{xy} & y 
  \end{array} \right) \qquad  \tilde{N}_m \in {\cal Z}^{(2)}_m \]
 The basis used here is $|{\cal C}^m_0> = \sqrt{x}|0^m> + \sqrt{y}|1^m>$ and $|{\cal C}^m_1> = 
 \sqrt{y}|0^m>-\sqrt{x}|1^m> $ 
                               
 The full configuration space is ${\cal Z} = \prod^\otimes_m\ {\cal Z}^{(2)}_m $  and the
 augmented Hamiltonian in $\Psi = {\cal Z}\otimes{\cal H}$ is :                             
 \begin{widetext}
\begin{eqnarray}
 \widetilde{H} & = & \sum_m \left(\epsilon_A \tilde{N}_m + \epsilon_B (\tilde{I}-\tilde{N}_m)\right)\otimes {\cal P}_m  + 
 \sum_{m}\sum_{m'} \left( t_{AA} \tilde{N}_m\otimes \tilde{N}_{m'}
+ t_{BB} (\tilde{I}-\tilde{N}_m)\otimes(\tilde{I}-\tilde{N}_m) + \right. \nonumber\\ 
& & t_{AB} \left\{\tilde{N}_m\otimes(\tilde{I}-\tilde{N}_{m'})+(\tilde{I}-\tilde{N}_m)\otimes\tilde{N}_{m'}\right\}]\otimes  {\cal T}_{mm'} 
\end{eqnarray}
\end{widetext}

The Augmented Space Theorem \cite{asr2} tells us that the configuration average of any function of  $\{n_m\}$ is :
\be  \ll F(\{n_m\})\gg \ = \  \langle G\otimes\{{\cal C}_0\}| \widetilde{F}(\{\tilde{N}_m\}) |G\otimes\{{\cal C}_0\}\rangle \ee

With $|\{{\cal C}_0\}\rangle = \prod^\otimes\ |{\cal C}^m_0\rangle $ and $\widetilde{F}\in {\cal Z}\otimes{\cal H} = \Psi$                                  

Now disorder can be combined with Recursion using the ideas of Augmented Space with all operators and 
states in $ \Psi\ =\ {\cal Z}\otimes{\cal H}$ 
It follows that :

\ber  
\ll S(\omega)\gg  =  \phantom{xxxxxxxxxxxxxxxxxxxxxxxxxxxxxx}&&\nonumber\\
 \int_0^\infty\ dt \exp(i\omega t)\left\langle G \otimes \{ {\cal C}_0\} \vert \tilde{j}(t)\  \tilde{j}(0)
 \vert G \otimes \{ {\cal C}_0\} \right\rangle &&\nonumber\\ 
 \eer

We calculate this using Generalized Recursion in the full augmented space $\Psi$. 
 The dielectric function  and optical conductivity are related :
 \begin{eqnarray}
  \ll\epsilon_2(\omega)\gg & = & 4\pi \lim_{\delta\rightarrow 0}\frac{\ll {\rm S}(z)\gg}{z^2} = \frac{\omega}{4\pi}\ll\sigma(\omega)\gg \nonumber\\
\end{eqnarray} 

\subsection{ The Optical Current.}

The calculations begin with a thorough electronic structure calculation using the tight-binding linear muffin-tin orbitals method. The technique, like many others, is based on the density functional theory, where the energetics depends entirely upon the charge and spin densities. However, the current operator which is characteristic of an excited electron : excited optically, electronically or magnetically, is described by transition probabilities which depend upon the wave-function. The wave-functions are expressed as linear combinations of the linearized basis functions : the muffin-tin orbitals. 

\begin{eqnarray}
 \Psi(\vec{r})  =  \sum_{RL} c_{RL} \left[\phi_{RL}(\vec{r}) + \sum_{RL,R'L'} h_{RL,R'L'}\ {\phi}'_{R'L'}(\vec{r}) \right]\nonumber\\
\end{eqnarray}

TB-LMTO-vLB-ASR calculations were done self-consistently and non-relativistically for a given geometry till the ``averaged relative error'' between the converged final and the previous iteration charge density and energy is reached. Nowhere do we use lattice translation symmetry so that Bloch's theorem plays no role.

{\par}The calculations begin with a thorough electronic structure calculation using the tight-binding linear muffin-tin orbitals method. The technique, like many others, is based on the density functional theory, where the energetics depends entirely upon the charge and spin densities. However, the current operator which is characteristic of an excited electron : excited optically, electronically or magnetically, is described by transition probabilities which depend upon the wave-function. The wave-functions are expressed as linear combinations of the linearized basis functions of the LMTO. The wave-function representation in the LMTO basis is :
\begin{eqnarray}
 \Psi(\vec{r}) & = & \sum_{RL} c_{RL} \left[\phi_{RL}(\vec{r})
 + \sum_{nL,n'L'} h_{nL,n'L'}\ {\phi}'_{n'L'}(\vec{r}) \right]\nonumber\\
\end{eqnarray}
where $\phi'_{nL}  =   \partial \phi(\vec{r},E)/\partial E $

The coefficients $c_{nL}$ are available from our TB-LMTO secular equation. To obtain the currents we follow the procedure of Hobbs {\etal}  \cite{hobs}.
 The optical current operator is given by :
 
\be j^\mu = \sum_{RL} J^\mu_{RL,RL} {\cal P}_{RL} + \sum_{RL}\sum_{R'L'} J^\mu_{RL.R'L'} {\cal T}_{RL,R'L'}\ee

\begin {eqnarray*}
 J^\mu_{RL,RL}\ & = & V^{(1),\mu}_{RL,RL} \\
J^\mu_{RL,R'L'}\ & = &  \sum_{R''L''} \left\{ V^{(2),\mu}_{RL,R''L''}h_{R''L'',R'L'} +\right. \ldots\\
 &+ & h_{RL,R''L''} V^{(3),\mu}_{R''L'',R'L'} \nonumber\\ 
& + & \left. \sum_{R''L''} V^{(4),\mu}_{R''L",R''',L'''}h_{R'''L''',R'L'} \right\} 
\end {eqnarray*}

where,

\begin{eqnarray*}
 V^{(1),\mu}_{RL,RL'}  =  \int_{r<s_R} d^3\vec{r}\ \phi^{*}_{RL'}(\vec{r}) (-i\nabla^\mu) 
                 \phi_{RL}(\vec{r}) \\ 
V^{(2),\mu}_{RL,RL'} = \int_{r<s_R} d^3\vec{r}\ \widehat{\phi}^{*}_{RL'}(\vec{r})  (-i\nabla^\mu) 
            \widehat{\phi}_{RL}(\vec{r}) \\                 
V^{(3),\mu}_{RL,RL'}  =  \int_{r<s_R} d^3\vec{r}\ \phi^{*}_{RL'}(\vec{r}) (-i\nabla^\mu) 
            \widehat{\phi}_{RL}(\vec{r}) \\
 V^{(4),\mu}_{RL,RL'} = \int_{r<s_R} d^3\vec{r}\ \widehat{\phi}^{*}_{RL'}(\vec{r}) (-i\nabla\mu) 
              \widehat{\phi}_{RL}(\vec{r}) 
\end{eqnarray*}

We have followed the prescription of Hobbs~\etal  \cite{hobs} to evaluate the integrals above. For details we again refer to reader to the above reference. We should note that we have so far not introduced the idea of reciprocal space. We expect our methodology to be applicable to situations w
The ASR method is a real-space technique for treating random disorder of alloys. Recursion  \cite{viswa,term} needs a countable basis $\{|n>\} \in Z^d $ which is provided by the TB-LMTO-ASA-vLB basis with $R$ labelling the atomic positions $\vec{R}_{n}$.

\end{document}